# QUANTUM DISCORD AND ENTANGLEMENT OF QUASI-WERNER STATES BASED ON BIPARTITE ENTANGLED COHERENT STATES

Manoj K Mishra[1], Ajay K Maurya[1] and Hari Prakash[1,2]

[1]Physics Department, University of Allahabad, Allahabad, India

[2]Indian Institute of Information Technology, Allahabad, India

email: manoj.qit@gmail.com, ajaymaurya.2010@rediffmail.com,

prakash_hari123@gmail.com, hariprakash@iiita.ac.in

**ABSTRACT:**

Present work is an attempt to compare quantum discord and quantum entanglement of Werner states formed with the four bipartite entangled coherent states (ECS) used recently for quantum teleportation of a qubit encoded in superposed coherent state. Out of these, two based on maximally ECS are exactly similar to the perfect Werner state, while the rest two, based on non-maximally ECS, which are called quasi-Werner states behave in somewhat different ways and their quantum discord and entanglement depends on the coherent amplitude and the measurement basis.

## I. INTRODUCTION

Quantum correlations have been comprehensively accepted as the main resource for different quantum information processing tasks. Einstein et al [1] and Schrödinger et al [2] at the outset introduced the concept of quantum correlation exhibited in a non-separable bipartite quantum state and this was called quantum entanglement (QE). Concurrence [3], Entanglement of formation [3], Tangle [4] and Negativity [5], defined by various researchers, quantify the entanglement existing in a bipartite or multipartite quantum state. Significant number of theoretical and experimental schemes [6-12] have been proposed for generating the entangled states based on photonic [6], atomic [7-9] and coherent state [10-12] qubits. Motivation behind all these works related to entanglement is due to its fundamental nature and its application in quantum information processing tasks like quantum computation [13], quantum teleportation [14], quantum dense coding [15], and quantum cryptography [16].

However, it has been discovered recently that QE is not the only type of quantum correlation and consequently the measures of QE like concurrence or entanglement of formation cannot be considered as a complete measure of quantum correlation. More explicitly, *non-zero entanglement guarantees the presence of quantum correlation, but zero*

*entanglement do not guarantees the absence of quantum correlation in a bipartite quantum state*. Zurek et al [17] proposed quantum discord (QD) as a measure of quantum correlation defined as, discrepancy between two corresponding expressions for classical mutual information, one obtained by using conditional entropy and the other by performing local measurement on any one of the subsystem. Classically the two mutual information are equivalent. Since QD is based on the total correlation (mutual information), it is capable to sense correlation in non-separable (entangled) as well as in separable states. It is well known that for pure bipartite quantum states, entanglement of formation and QD are equal, while for mixed separable states QD can take non-zero value. Appreciable amount of works related to theoretical development of QD [18], its dynamical property under the effect of decoherence [19] and its comparison with QE have been done over the past decade [20]. Boixo et al. [21] proposed a protocol known as quantum locking of classical correlations and shown that QD allows us to have negligible size key for an encrypted classical message, while with classical resources we must have key of size comparable to message to approach unconditional security. QD plays an important role in investigating the power of schemes for quantum computation without QE [22]. Lei Wang et al [23] shown how to teleport a single qubit state using only non-zero QD and state tomography regardless of if there is QE or not. Thus QD not only explores the quantum correlations, but also promises for quantum information processing applicability.

Superiority of QD over QE in sensing the quantum correlation and its applicability attracted researchers to investigate the dynamics of QD in different quantum mechanical systems, like bipartite two-level atomic system interacting with a cavity field [24], quantum dots [25], spin chains [26] and in different quantum states like bipartite Bell diagonal state, bipartite X class states [27] and multi partite states [28]. These studies are important as they can help us in retrieving the quantum properties of the system as well as can act as tool for development of QD theory. Usually the system is constructed of superposition of orthogonal states. For instance the Werner state based on standard Bell basis uses states like $|0\rangle, |1\rangle$, and of course properties of QD of these states are well known. Our interest here is to see what kind of properties of QD will appear if we have superposition of non-orthogonal states. One such system is bipartite superposition of Glauber coherent states of radiation that are $\pi$ radian out of phase also called entangled coherent states (ECS) [29]. The QE [30] and statistical properties [31] of ECS have already been studied by many authors. ECS serves as a resource in universal quantum computation, teleportation and for quantum networks. Since ECS are pure states, there will not be any difference between QD and Entanglement of formation. In this paper, we consider the quasi-Werner mixed states based on ECS and study how non-orthogonality of the basis influences OD and entanglement.

## II. QUANTUM DISCORD AND ENTANGLEMENT

Consider two random variables *X* and *Y*. The classical mutual information (total classical correlation) between the variables is given by

$$I(X:Y) = H(X) + H(Y) - H(X,Y), \qquad (1)$$

where $H(X) = -\sum_{x \in X} p(x) \log p(x)$ is the Shannon entropy that gives the uncertainty about the random variable *X*, and $H(X,Y) = -\sum_{x \in X, y \in Y} p(x,y) \log p(x,y)$ is the joint entropy. All the probability distributions are calculated from the joint probability $p(X,Y)$:

$$p(X) = \sum_y p(X, Y=y); \quad p(Y) = \sum_x p(X=x, Y).$$

The conditional entropy measures how much uncertainty is left, on average, regarding the value of *X*, for given value of *Y*. It can be written as

$$H(X|Y) = \sum_{y \in Y} p(y) H(X|Y=y) = -\sum_{y \in Y} p(y) \sum_{x \in X} p(x|y) \log p(x|y).$$

Using Bayes rule, $p(X|Y=y) = p(X, Y=y)/p(Y=y)$, above definition of conditional entropy reduces to

$$H(X|Y) = H(X,Y) - H(Y). \qquad (2)$$

Using equation (2) in equation (1) we get another expression for classical mutual information,

$$J(X:Y) = H(X) - H(X|Y). \qquad (3)$$

Zurek et al [17] generalized the two equivalent expressions (1) and (3) of the classical mutual information to quantum systems. Replacing the probability distributions with density matrices and the Shannon entropy with the Von Neumann entropy $S(\rho) = -Tr(\rho \log \rho)$ in equation (1), one has

$$I(X:Y) = S(\rho_X) + S(\rho_Y) - S(\rho_{X,Y}), \qquad (4)$$

where $\rho_X$ represents the reduced density matrix.

In quantum theory, conditional entropy can be obtained by applying measurement over system *Y*. Zurek et al [17] assumed a perfect measurements of *Y* defined by a set of one-dimensional projectors $\{\Pi_j^Y\}$, such that $\sum_j \Pi_j^Y = 1$. The state of *X* after this measurement is given by, $\rho_{X|\Pi_j^Y} = (\Pi_j^Y \rho_{XY} \Pi_j^Y)/Tr(\Pi_j^Y \rho_{XY})$ with probability $p_j = Tr(\Pi_j^Y \rho_{XY})$. Then, the conditional entropy is defined as $S(\rho_{X|\{\Pi_j^Y\}}) = \sum_j p_j S(\rho_{X|\Pi_j^Y})$ with $S(\rho_{X|\Pi_j^Y}) = -Tr(\rho_{X|\Pi_j^Y} \log \rho_{X|\Pi_j^Y})$ and the quantum equation corresponding to equation (3) can be written as

$$J(X:Y)_{\{\Pi_j^Y\}} = S(\rho_X) - S(\rho_{X|\{\Pi_j^Y\}}). \tag{5}$$

The two classically equivalent expressions (1) and (3) for mutual information, lead to quantum mechanical expressions (4) and (5), respectively. Expressions (4) and (5) are not identical and QD is defined as the difference,

$$D(X:Y)_{\{\Pi_j^Y\}} = I(X:Y) - J(X:Y)_{\{\Pi_j^Y\}} = S(\rho_Y) - S(\rho_{X,Y}) + S(\rho_{X|\{\Pi_j^Y\}}), \tag{6}$$

and the minima of this is defined as

$$\delta(X:Y)_{\{\Pi_j^Y\}} = \min_{\{\Pi_j^Y\}}[D(X:Y)_{\{\Pi_j^Y\}}]. \tag{7}$$

Zurek et al [17], plotted $D$ for state, $\rho_{XY} = \frac{1}{2}[|00\rangle\langle 00| + |11\rangle\langle 11|] + \frac{a}{2}[|00\rangle\langle 11| + |11\rangle\langle 00|]$, with $0 \leq a \leq 1$, for various values of $a$ and measurement parameter $\theta$, evaluated for measurement basis state $\{\cos\theta|0\rangle + e^{i\phi}\sin\theta|1\rangle, e^{-i\phi}\sin\theta|0\rangle - \cos\theta|1\rangle\}$ with $\phi = 1\,rad$. The values of $a$ and $\theta$ in their plot [17] ran from 0 to 1 and from $-\pi$ to $\pi$ respectively. The plot showed minimum value of $D$ at $a=0$, $\theta=0$ and the authors remarked that only in the case of complete einselection $a=0$, there exists a basis for which discord disappears. We repeated the calculations for this problem and obtained the expression for QD in the form,

$$D = 1 + [\frac{1+a}{2}\log\frac{1+a}{2} + \frac{1-a}{2}\log\frac{1-a}{2}] - [\frac{1+[1-(1-a^2)\sin^2 2\theta]^{1/2}}{2}$$
$$\times \log\frac{1+[1-(1-a^2)\sin^2 2\theta]^{1/2}}{2} + \frac{1-[1-(1-a^2)\sin^2 2\theta]^{1/2}}{2} \tag{8}$$
$$\times \log\frac{1-[1-(1-a^2)\sin^2 2\theta]^{1/2}}{2}].$$

This is evident by the fact that eigenvalues of $\rho_Y$ and $\rho_{XY}$ are, $(\pm 1/2)$ and $(0,0,[1\pm a]/2)$ respectively, and eigenvalues of $\rho_{X|\Pi_j^Y}$ for $j=1, 2$ are $(0,0,[1\pm(1-(1-a^2)\sin^2 2\theta)^{1/2}]/2)$.

We plot $D$ given by equation (8) in Fig.1, which shows that $D$ is periodic in measurement parameter $\theta$ with minima at $a=0$, $\theta = 0, \pi/2, \pi$, while in Ref. [17] only one minima has been shown in the domain $-\pi \leq \theta \leq \pi$. Also it must be noted that Zurek et al [17], used $\phi = 1\,rad$, while from equation (8) it is clear that the value of $D$ does not depends on the phase angle $\phi$ whatever be the value of $\theta$. Although this agree with the same remark of Zurek et al [17] that only in the case of complete einselection $a = 0$ there exists a basis for which QD disappears, but the correct expression and plot of $D$ for this particular state are those given by equation (8) and Fig.1, respectively. This becomes clear also by putting $a = 1$ which gives, from equation (8), $D = 1$, as is shown in Fig. 1, while Zurek et al [17] show $D \approx 7$ in their Fig.1.

Concurence and entanglement of formation are the two frequently used measures of QE for a bipartite state. Wootters et al [3] defined concurrence as,

$$C = \max\{0,(\lambda_1 - \lambda_2 - \lambda_3 - \lambda_4)\}, \tag{9}$$

where $\lambda_1, \lambda_2, \lambda_3, \lambda_4$ are the non-negative square roots of eigenvalues of the matrix $(\rho\tilde{\rho})_{XY}$, with $\tilde{\rho}_{XY} = \sigma_y \otimes \sigma_y (\rho^*_{XY})\sigma_y \otimes \sigma_y$, in decreasing order. Entanglement of formation is defined as a monotonic function of concurrence by the relation,

$$E = H(x) = -x\log x - (1-x)\log(1-x), \tag{10}$$

where $H$ is the Shannon entropy function and $x = (1+\sqrt{1-C^2})/2$.

### III. QUANTUM DISCORD AND ENTANGLEMENT OF QUASI-WERNER STATES BASED ON ECS

The coherent state of radiation, defined as the eigenstate of annihilation operator $a$,

$$a|\alpha\rangle = \alpha|\alpha\rangle, \quad |\alpha\rangle = e^{-\frac{1}{2}|\alpha|^2} \sum_{n=0}^{\infty} \frac{\alpha^n}{\sqrt{n!}}|n\rangle \tag{11}$$

is used to encode a qubit when superposed with $|-\alpha\rangle$. Since $|\pm\alpha\rangle$ are not orthogonal, a orthogonal basis is obtained by considering odd and even coherent state $|\pm\rangle$ given by

$$|\pm\rangle = N_\pm[|\alpha\rangle \pm |-\alpha\rangle], \quad N_\pm = [2(1 \pm x^2)]^{-1/2}, \quad x = e^{-|\alpha|^2}, \tag{12}$$

Consider four bipartite entangled coherent state

$$|\psi^\pm\rangle_{XY} = n_\pm[|\alpha,\alpha\rangle \pm |-\alpha,-\alpha\rangle]_{XY}, \quad |\phi^\pm\rangle_{XY} = n_\pm[|\alpha,-\alpha\rangle \pm |-\alpha,\alpha\rangle]_{XY}, \tag{13}$$

where,

$$n_\pm = [2(1 \pm x^4)]^{-1/2}. \tag{14}$$

If we express $|\psi^\pm\rangle_{XY}$ and $|\phi^\pm\rangle_{XY}$ in the orthogonal basis of states $|\pm\rangle$ we can write

$$|\psi^+\rangle_{XY} = \frac{n_+}{2}[\frac{|+,+\rangle}{N_+^2} + \frac{|-,-\rangle}{N_-^2}]_{XY}, \quad |\phi^+\rangle_{XY} = \frac{n_+}{2}[\frac{|+,+\rangle}{N_+^2} - \frac{|-,-\rangle}{N_-^2}]_{XY}, \quad (15.a)$$

$$|\psi^-\rangle_{XY} = \frac{[|+,-\rangle + |-,+\rangle]}{\sqrt{2}}, \quad |\phi^-\rangle_{XY} = \frac{[|-,+\rangle - |+,-\rangle]}{\sqrt{2}}. \quad (15.b)$$

It is easily seen that the two states $|\psi^-\rangle$ and $|\phi^-\rangle$ are maximally entangled states with unit concurrence, while the states $|\psi^+\rangle$ and $|\phi^+\rangle$ are non-maximally entangled with concurrence $C = (1-x^4)(1+x^4)^{-1}$ and are non-orthogonal mutually. However, these become almost maximally entangled and almost orthogonal for appreciable mean photon numbers. These four states (equation 15) in the limit of large mean photon number form a complete orthogonal basis just like standard Bell states. Werner mixed states are given by $\rho(\psi, a) = (1-a)\frac{I}{4} + a|\psi\rangle\langle\psi|$, where $|\psi\rangle$ is the maximally entangled Bell state. If we use ECS, instead of Bell states, we get

$$\rho(\psi^-, a) = (1-a)\frac{I}{4} + a|\psi^-\rangle\langle\psi^-|, \quad (16.a)$$

$$\rho(\phi^-, a) = (1-a)\frac{I}{4} + a|\phi^-\rangle\langle\phi^-|, \quad (16.b)$$

$$\rho(\psi^+, a) = (1-a)\frac{I}{4} + a|\psi^+\rangle\langle\psi^+|, \quad (16.c)$$

$$\rho(\phi^+, a) = (1-a)\frac{I}{4} + a|\phi^+\rangle\langle\phi^+|. \quad (16.d)$$

Since the states in equation (15.b) are maximally entangled states, therefore, $\rho(\psi^-, a)$ and $\rho(\phi^-, a)$ are equivalent to perfect Werner mixed states. The rest two states are non-maximally entangled and therefore, states $\rho(\psi^+, a)$ and $\rho(\phi^+, a)$ are known as quasi-Werner states. We are interested in exploring difference in dynamics of discord of quasi-Werner states to that of perfect Werner states. Density matrix $\rho(\psi^+, a)$ is given by,

$$\rho_{XY}(\psi^+,a) = \begin{pmatrix} \frac{1}{4}+\frac{a}{4}\left(\frac{n_+^2}{N_+^4}-1\right) & 0 & 0 & \frac{an_+^2}{4N_+^2 N_-^2} \\ 0 & \frac{1-a}{4} & 0 & 0 \\ 0 & 0 & \frac{1-a}{4} & 0 \\ \frac{an_+^2}{4N_+^2 N_-^2} & 0 & 0 & \frac{1}{4}+\frac{a}{4}\left(\frac{n_+^2}{N_-^4}-1\right) \end{pmatrix}. \tag{17}$$

Eigenvalues of this matrix are,

$$\{(1-a)/4, (1-a)/4, (1-a)/4, (1+3a)/4\}. \tag{18}$$

Eigen-values of reduced density matrix $\rho_Y(\psi^+,a)$ are,

$$\{(1-a)/2 + (an_+^2)/(4N_+^4), (1-a)/2 + (an_+^2)/(4N_-^4)\}. \tag{19}$$

To calculate conditional entropy, we perform complete measurement of quantum mode $Y$ defined by a set of one-dimensional projectors $\{\Pi_j^Y\} \equiv \{|\pi_0\rangle\langle\pi_0|, |\pi_1\rangle\langle\pi_1|\}$ where $|\pi_0\rangle = \cos\theta|+\rangle + e^{i\phi}\sin\theta|-\rangle$, $|\pi_1\rangle = \sin\theta|+\rangle - e^{i\phi}\cos\theta|-\rangle$. State of mode $X$ after outcome corresponding to $\Pi_0^Y$ and $\Pi_1^Y$ are

$$\rho_{X|\Pi_0^Y} = \frac{1}{P_0}\begin{pmatrix} \frac{1}{4}+\frac{a}{4}\left(\frac{n_+^2}{N_+^4}\cos^2\theta-1\right) & \frac{an_+^2}{4N_+^2 N_-^2}e^{i\phi}\sin\theta\cos\theta \\ \frac{an_+^2}{4N_+^2 N_-^2}e^{-i\phi}\sin\theta\cos\theta & \frac{1}{4}+\frac{a}{4}\left(\frac{n_+^2}{N_-^4}\sin^2\theta-1\right) \end{pmatrix} \tag{20}$$

and

$$\rho_{X|\Pi_1^Y} = \frac{1}{P_1}\begin{pmatrix} \frac{1}{4}+\frac{a}{4}\left(\frac{n_+^2}{N_+^4}\sin^2\theta-1\right) & \frac{-an_+^2}{4N_+^2 N_-^2}e^{i\phi}\sin\theta\cos\theta \\ \frac{-an_+^2}{4N_+^2 N_-^2}e^{-i\phi}\sin\theta\cos\theta & \frac{1}{4}+\frac{a}{4}\left(\frac{n_+^2}{N_-^4}\cos^2\theta-1\right) \end{pmatrix}, \tag{21}$$

respectively, with probability $P_0$ and $P_1$ given by

$$P_0 = \frac{1-a}{2} + \frac{an_+^2}{4}\left(\frac{\cos^2\theta}{N_+^4} + \frac{\sin^2\theta}{N_-^4}\right), P_1 = \frac{1-a}{2} + \frac{an_+^2}{4}\left(\frac{\cos^2\theta}{N_-^4} + \frac{\sin^2\theta}{N_+^4}\right) \tag{22}$$

The eigen-values of $\rho_{X|\Pi_0^Y}$ and $\rho_{X|\Pi_1^Y}$ are respectively,

$$\{(1-a)/4P_0, 1-(1-a)/4P_0\} \tag{23}$$

and

$$\{(1-a)/4P_1, 1-(1-a)/4P_1\}. \tag{24}$$

Using equation (18, 19, 23, 24 and 6), QD for quasi-Werner state $\rho(\psi^+,a)$ is given as

$$D(X:Y)_{\{\Pi_j^Y\}} = -[(1-a)/2 + (an_+^2)/(4N_+^4)]\log[(1-a)/2 + (an_+^2)/(4N_+^4)]$$
$$-[(1-a)/2 + (an_-^2)/(4N_-^4)]\log[(1-a)/2 + (an_+^2)/(4N_-^4)]$$
$$+[3(1-a)/4]\log[(1-a)/4] + [(1+3a)/4]\log[(1+3a)/4] \quad (25)$$
$$-\sum_{j=0,1} P_j[\{(1-a)/4P_j\}\log\{(1-a)/4P_j\}]$$
$$-\sum_{j=0,1} P_j[\{1-(1-a)/4P_j\}\log\{1-(1-a)/4P_j\}]$$

where $P_{0,1}$ are defined by equation (22). One can verify that QD for quasi-Werner state $\rho(\phi^+,a)$ is equal to that for $\rho(\psi^+,a)$. It should be noted that in this paper, log is taken at the base 2. Following the same procedure, quantum discord for Werner states, $\rho(\psi^-,a)$, $\rho(\phi^-,a)$ is given by

$$D'(X:Y)_{\{\Pi_j^Y\}} = -1 + [3(1-a)/4]\log[(1-a)/4] + [(1+3a)/4]\log[(1+3a)/4]$$
$$-[(1-a)/2]\log[(1-a)/2] - [(1+a)/2]\log[(1+a)/2]. \quad (26)$$

From equations (25, 26), it is clear that QD of quasi-Werner states, $D$ given by equation (25), depends upon the mixing parameter '$a$', measurement parameter $\theta$ and mean photon numbers, while QD of perfect Werner state $D'$ given by equation (26) is independent of measurement parameter and mean photon numbers. Fig. 2 shows the variation of $D(X:Y)_{\{\Pi_j^Y\}}$ with '$a$' and $\theta$, for different values of mean photon numbers. Fig. 3 shows the variation of $D'(X:Y)_{\{\Pi_j^Y\}}$ with '$a$'. From Fig.2 it is clear that QD of quasi-Werner states increases as mean photon number increases and for appreciable mean photon numbers, it becomes independent of measurement basis and behaves almost similar to that for perfect Werner state (Fig. 3). However, for small mean photon number, $D(X:Y)_{\{\Pi_j^Y\}}$ depends effectively on measurement basis with minimum at $\theta = 0, \pi/2, \pi$.

It is very interesting to see that for very small mean photon number $|\alpha|^2 \leq 0.02$, $D(X:Y)_{\{\Pi_j^Y\}}$ show different nature against both '$a$' and $\theta$ as compared to that for $|\alpha|^2 > 0.02$. For $|\alpha|^2 \leq 0.02$ and $0 < \theta < \pi/2$, $D(X:Y)_{\{\Pi_j^Y\}}$ first increases with '$a$' and then decreases almost to zero at $a=1$. For $|\alpha|^2 > 0.02$ and $0 < \theta < \pi/2$, the value of $D(X:Y)_{\{\Pi_j^Y\}}$ first increases and then decreases to some appreciable nonzero value at $a=1$. For $|\alpha|^2 > 0.2$ and $0 < \theta < \pi/2$, value of $D(X:Y)_{\{\Pi_j^Y\}}$ takes general increasing nature with increasing 'a' which is similar to that for $\theta = 0, \pi/2$ for which minimum of quantum discord occurs.

Using $\theta = \pi/2$, in equation (25), we plotted minimum QD of quasi-Werner state, $\delta(X:Y)_{\{\Pi_j^Y\}} = \min_{\{\Pi_j^Y\}}[D(X:Y)_{\{\Pi_j^Y\}}]$ in Fig. 4, from where it is clear that minimum QD, $\delta(X:Y)_{\{\Pi_j^Y\}}$ of quasi-Werner states also depends on mean photon number and increases with increasing the value of mean photon number.

Eigen-values of matrix $(\rho\tilde{\rho})_{X,Y}$ for quasi-Werner states $\rho(\psi^+, a)$ and $\rho(\phi^+, a)$ are same and given by

$$[(1-a)/4, (1-a)/4, \{((1-a)/4 + (an_+^2/4N_+^4))((1-a)/4 + (an_+^2/4N_-^4))\}^{-1/2} \pm (an_+^2/4N_+^2N_-^2)] \tag{27}$$

Eigen-values of matrix $(\rho\tilde{\rho})_{X,Y}$ for perfect Werner states $\rho(\psi^-, a)$ and $\rho(\phi^-, a)$ are same, given by

$$[(1-a)/4, (1-a)/4, (1-a)/4, (1+3a)/4]. \tag{28}$$

Using equations (9-10) and (27-28), we have plotted the entanglement of formation ($E$) and Difference between QD and entanglement of formation ($\delta - E$) for quasi-Werner state and Perfect Werner state in Fig.4 and Fig.3 respectively.

## IV. DISCUSSION AND SUMMARY

In section III, we considered Werner forms based on ECS. Use of maximally ECS instead of Bell states gives perfect Werner state for which QD depends upon the mixing parameter $a$ and not on measurement basis. This is evident by the fact that perfect Werner states are invariant under local operations. From Fig. 3, it is clear that for $a > 1/3$, entanglement of formation increases with higher rate than QD, which in turn implies that as the state reaches to less mixed nature the quantum correlation increases, but contributes more to the entanglement than the quantum discord.

Use of non-maximally ECS instead of Bell states gives quasi-Werner state for which QD depends on mixing parameter $a$, measurement basis and mean photon numbers. For small mean photon numbers it is found that QD is small and more sensitive to measurement basis, it disappears at $\theta = 0, \pi/2, \pi$. As mean photon number increases, the QD increases and becomes less sensitive to measurement basis. Fig.4 shows that maximum difference between QD and entanglement of formation for large mean photon number occurs in the range

$0.4 < a < 0.5$, while for small mean photon numbers this range shifts to slightly higher values of $a$.

In section II, we pointed out a computational mistake in Ref. [17] about the QD of the state $\rho_{XY} = \frac{1}{2}[|00\rangle\langle 00| + |11\rangle\langle 11|] + \frac{a}{2}[|00\rangle\langle 11| + |11\rangle\langle 00|]$, for which it was reported that QD, in case of complete einselection, i.e., $a = 0$, disappears at $\theta = 0$ and has nonzero values at $\theta = 0, \pi/2, \pi$ [see Fig.1 in Ref. 17]. However we revaluated this and found that QD, in case of complete einselection i.e., $a = 0$ disappears at $\theta = 0, \pi/2, \pi$. Also from equation (8) it is clear that QD does not depends on the phase angle of measurement basis, while in Ref. [17] $\phi = 1\,rad.$ is used.

In summary we studied the quantum discord and quantum entanglement of Werner states formed with the four bipartite entangled coherent states (ECS) used recently for quantum teleportation of a qubit encoded in superposed coherent state. It is found that, if $|\alpha|^2$ is not too small, for both Werner state and quasi-Werner states, QD and entanglement of formation increase as mixing parameter $a$ is increased (i.e., when mixedness of quantum state decreases), but this increase in quantum correlation contributes more to quantum entanglement than to QD. QD of perfect Werner states due to its invariant nature under local operation is independent of measurement basis, while for quasi-Werner states it depends upon measurement basis as well as on mean photon number. However, for large mean photon numbers since quasi-Werner states tends to perfect Werner state, therefore dependence of QD on the measurement basis disappears.

**Acknowledgements:**

We are grateful to Prof. N. Chandra and Prof. R. Prakash for their interest and stimulating discussions. Discussions with Ajay Kumar Yadav and Vikram Verma are gratefully acknowledged. One of the authors (MKM) acknowledges the UGC for financial support under UGC-SRF fellowship scheme.

**References:**

[1] A. Einstein, B. Podolsky and N. Rosen, Phys. Rev. 47, 777 (1935).

[2] E. Schr¨odinger, Naturwissenschaften 23, 807 (1935).

[3] W. K. Wootters, Phys. Rev. Lett. 80, 2245 (1998); C. H. Bennett, D. P. DiVincenzo, J. Smolin, and W. K. Wootters, Phys. Rev. A 54, 3824 (1996); S. Hill and W. K. Wootters,

Phys. Rev. Lett. 78, 5022 (1997); P. Rungta, V. Buzek, C. M. Caves, M. Hillery, and G. J. Milburn, Phys. Rev. A 64, 042315.

[4] V. Coffman, J. Kundu and W. K. Wootters, Phys. Rev. A 61, 052306 (2000).

[5] K. Zyczkowski, P. Horodecki, A. Sanpera and M. Lewenstein, Phys. Rev. A 58, 883 (1998); G. Vidal and R. F. Werner, Phys. Rev. A 65, 032314 (2002); M. B. Plenio, Phys. Rev. Lett. 95, 090503 (2005).

[6] P. G. Kwiat et al., Phys. Rev. Lett. 75, 4337 (1995); Y. H. Shih, C. O. Alley, Phys. Rev. Lett. 61, 2921 (1988); Y. H. Shih, A. V. Sergienko, M. H. Rubin, T. E. Kiess, C. O. Alley, Phys. Rev. A 50, 23 (1994); D. Bouwmeester, J. W. Pan, K. Mattle, M. Eibl, H. Weinfurter and A. Zeilinger, Nature 390, 575 (1997).

[7] E. Hangley et al., Phys. Rev. Lett. 79, 1 (1997).

[8] S. B. Zheng and G. C. Guo, Phys. Rev. Lett. 85, 2392 (2000).

[9] S. Lloyd, M. S. Shahriar and J. H. Shapiro, Phys. Rev. Lett. 87, 167903 (2001).

[10] C. C. Gerry, Phys. Rev. A 59, 4095 (1999).

[11] J. C. Howell and J. A. Yeazell, Phys. Rev. A 62, 012102 (2000).

[12] J. Q. Liao and L. M. Kuang, J. Phys. B: At. Mol. Opt. Phys. 40, 1845 (2007).

[13] D. P. DiVincenzo, Science 270, 255 (1995); D. Gottesman and I. L. Chuang, Nature (London) 402, 390 (1999); P. W. Shor, Proceedings of the Symposium on the Foundationsof Computer Science, 1994, Los Alamitos, California (IEEE Computer Society Press, New York, 1994), pp. 124–134; L. K. Grover, Phys. Rev. Lett. 79, 325 (1996).

[14] C. H. Bennett, G. Brassard, C. Crepeau, R. Jozsa, A. Peres, and W. K. Wootters, Phys. Rev. Lett. 70, 1895 (1993); D. Bouwmeester, J. W. Pan, K. Mattle, M. Eibl, H. Weinfurter, and A. Zeilinger, Nature (London) 390, 575 (1997); D. Boschi, S. Branca, F. De Martini, L. Hardy, and S. Popescu, Phys. Rev. Lett. 80, 1121 (1998).

[15] C. H. Bennett and S. J. Wiesner, Phys. Rev. Lett. 69, 2881 (1992); K. Mattle, H. Weinfurter, P. G. Kwiat, and A. Zeilinger, ibid. 76, 4656 (1996); S. L. Braunstein and H. J. Kimble, Phys. Rev. A 61, 042302 ~2000.

[16] C. H. Bennett, G. Brassard, and N. D. Mermin, Phys. Rev. Lett. 68, 557 (1992).

[17] H. Ollivier and W. H. Zurek, Phys. Rev. Lett. 88, 017901 (2001); W. H. Zurek, Rev. Mod. Phys. 75, 715 (2003).

[18] L. Henderson and V. Vedral, J. Phys. A: Math. Gen. 34, 6899 (2001); J. Oppenheim et al, Phys. Rev. Lett. 89, 180802 (2002); V. Vedral, Phys. Rev. Lett. 90, 050401(2003); M. Horodecki et al, Phys. Rev. A 71, 062307 (2005); M. Piani, P. Horodecki and R.

Horodecki, Phys. Rev. Lett. 100, 090502 (2008); S. Luo, Phys. Rev. A 77 042303 (2008); J. Maziero, L. C. Cel´eri and R. M. Serra arXiv:1004.2082[quant-ph] (2010); A. Datta arXiv:1003.5256 [quant-ph] (2010); P. Giorda and M. G. A. Paris, Phys. Rev. Lett. 105, 020503 (2010); R. Vasile, P. Giorda, S. Olivares, M. G. A. Paris and S. Maniscalco, arXiv:1005.1043 [quant-ph] (2010); G. Adesso and A. Datta, Phys. Rev. Lett. 105, 030501 (2010); K. Modi, T. Paterek, W. Son, V. Vedral and M. Williamson, Phys. Rev. Lett. 104, 080501 (2010); B. Daki´c, V. Vedral and C. Brukner, arXiv:1004.0190 [quant-ph] (2010); D. Cavalcanti, L. Aolita, S. Boixo, K. Modi, M. Piani and A. Winter, arXiv:1008.3205 [quant-ph] (2010); M. Ali, A. R. P. Rau and G. Alber, Phys. Rev. A 81, 042105 (2010); V. Madhok and A. Datta, arXiv:1008.4135 [quant-ph] (2010); D. Girolami, M. Paternostro and G. Adesso, arXiv:1008.4136 [quant-ph] (2010).

[19] J. Maziero, L. C. Cel´eri, R. M. Serra and V. Vedral, Phys. Rev. A 80, 044102 (2009); T. Werlang, S. Souza, F. F. Fanchini and C. J. Villas-Boas, Phys. Rev. A 80, 024103 (2009); F. F. Fanchini et al, Phys. Rev. A 81, 052107 (2010); J. Maziero et al, Phys. Rev. A 81, 022116 (2010); J. Paz and A. J. Roncaglia, Phys. Rev. A 80, 042111 (2009); L. Mazzola, J. Piilo and S. Maniscalco, Phys. Rev. Lett. 104, 200401(2010); J. S. Xu, X. Y. Xu, C. F. Li, C. J. Zhang, X. B. Zou and G. C. Guo, Nat. Commun. 1 7 (2010).

[20] T. Yu and J. H. Eberly, Phys. Rev. Lett. 93, 140404 (2004); M. Yonac, T. Yu and J. H. Eberly, J. Phys. B: At. Mol. Opt. Phys. 39, S621 (2006).

[21] S. Boixo, L. Aolita, D. Cavalcanti, K. Modi and A. Winter, arXiv:1105.2768v2 [quant-ph] (2011).

[22] A. Datta, A. Shaji and C. Caves, Phys. Rev. Lett. 100, 050502 (2008); A. Datta, e-print arXiv:0807.4490; A. Datta and S. Gharibian, Phys. Rev. A 79, 042325 (2009).

[23] Lei Wang, Jie-Hui Huang, Jonathan P. Dowling and Shi-Yao Zhu, arXiv: 1106.5097 [quant-ph] (2011).

[24] D. Z. Rossatto, T. Werlang, E. I. Duzzioni, and C. J. Villas-Boas1, Phys. Rev. lett. 107, 153601 (2011) ; C. Z. Wang, C. X. Li, L. Y. Nie and J. F. Li, J. Phys. B: At. Mol. Opt. Phys. 44, 015503 (2011); H. T. Dung, A. S. Shumovsky, and N. N. Bogolubov, Opt. Commun. 90, 322 (1992); Q. Yi, X. U. Jing-Bo, Chin. Phys. Lett. 28, 070306 (2011).

[25] F. F. Fanchini, L. K. Castelano and A. O. Caldeira, New J. Phys. 12, 073009 (2010).

[26] T. Werlang and G. Rigolin, Phys. Rev. A 81, 044101 (2010); T. Werlang, C. Trippe, G. A. P. Ribeiro and G. Rigolin, Phys. Rev. Lett. 105, 095702 (2010); J. Maziero, H. C. Guzman, L. C. Celeri, M. S. Sarandy and R. M. Serra, arXiv:1002.3906 [quant-ph] (2010).

[27] D. L. Matthias and M. C. Caves, Phys. Rev. Lett. 105, 150501 (2010);S. Luo, Phys. Rev. A 77, 042303 (2008); M. Ali, A. R. P. Rau, and G. Alber, Phys. Rev. A 81, 042105 (2010); Bo Li, Zhi-Xi Wang, and Shao-Ming Fei, Phys. Rev. A 83, 022321 (2011).

[28] C. C. Rulli and M. S. Sarandy, Phys. Rev. A 84, 042109 (2011).

[29] B. C. Sanders, Phys. Rev. A 45, 6811 (1992).

[30] X. Wang and B.C. Sanders, Phys. Rev. A 65, 012303 (2001); S. Sivakumar, Int. J. Theor. Phys. 48, 894 (2008); O. Hirota and S.J. Van Enk, quant-ph/0101096 (2001).

[31] C-lin Chai, Phys. Rev. A 46, 7187 (1994); P. Tombesi and A. Mecozzi, J. Opt. Soc. Am. B 4, 1700 (1987).

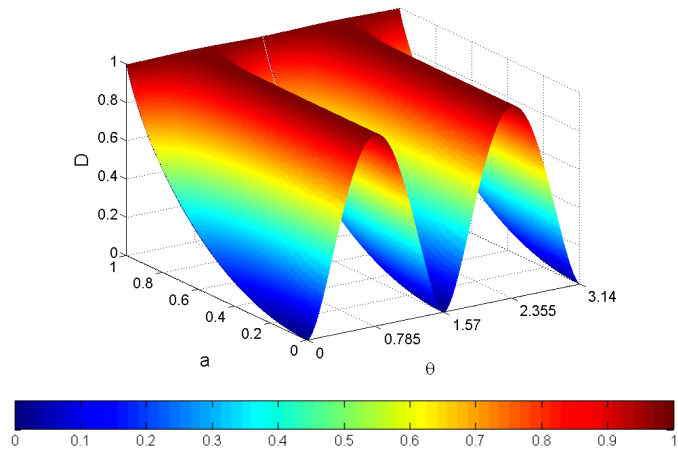

Fig.1: Variation of $D$ (Eq. 8) with respect to '$a$' and measurement parameter $\theta$.

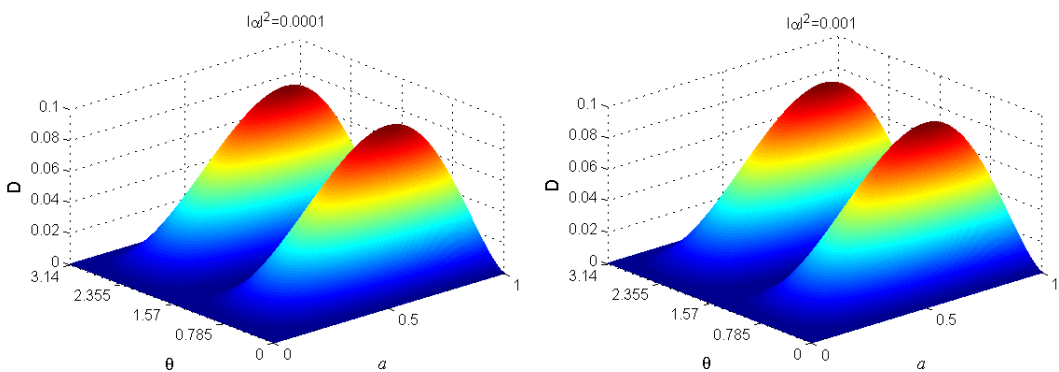

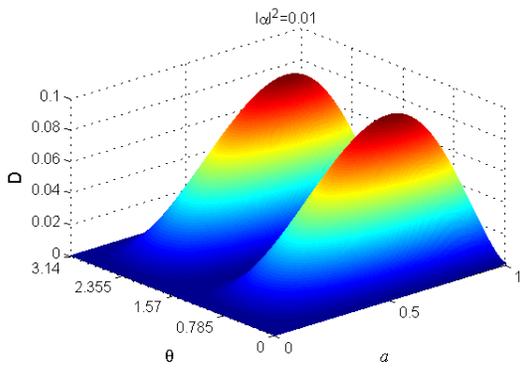
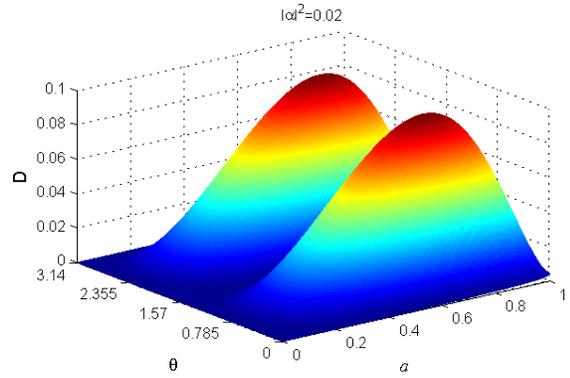
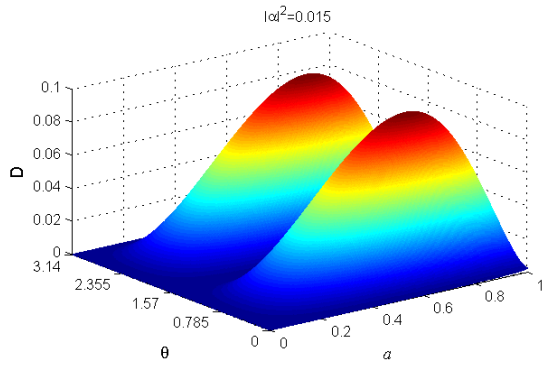
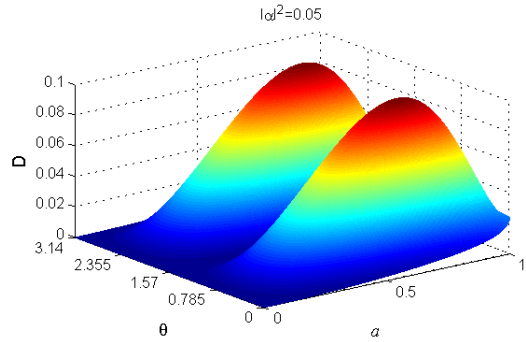
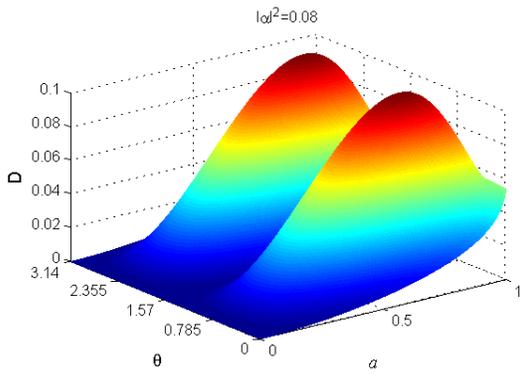
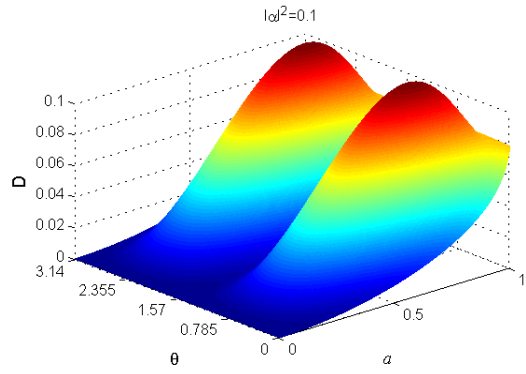
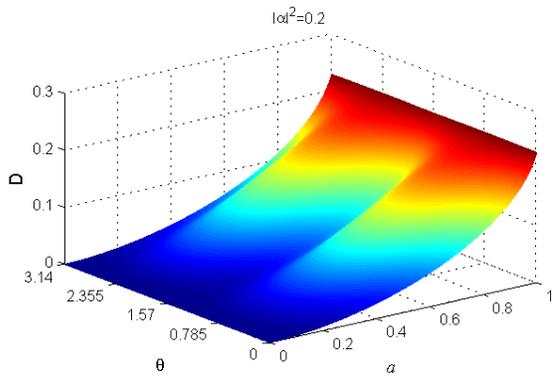
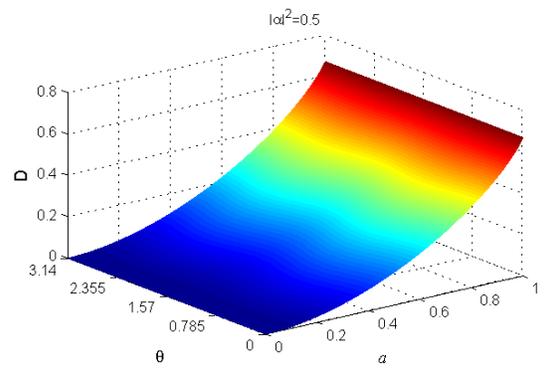

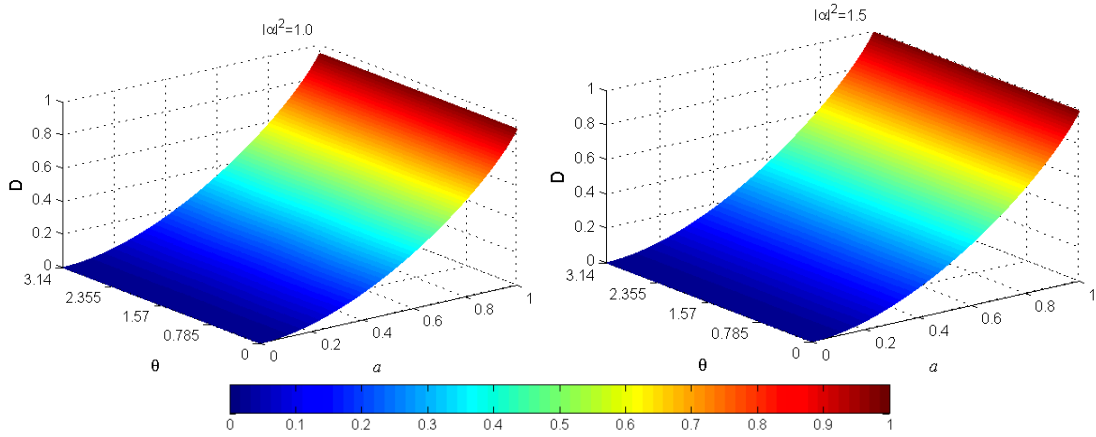

Fig.2. Variation of quantum discord (*D*) of quasi-Werner states with respect to measurement parameter ($\theta$) and mixing parameter (*a*) for different values of mean photon number ($|\alpha|^2$)

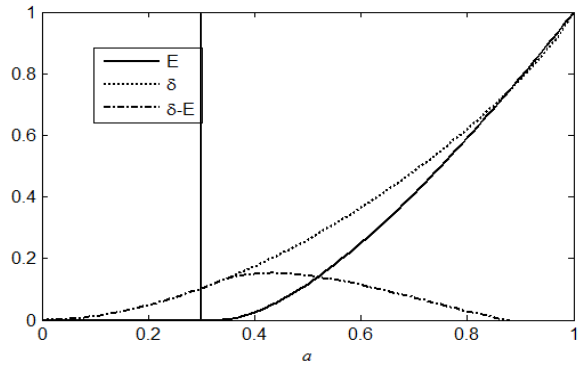

Fig.3. Variation of Entanglement of formation (*E*), minimum discord ($\delta$ or *D'*) and their difference ($\delta - E$) for Perfect Werner states.

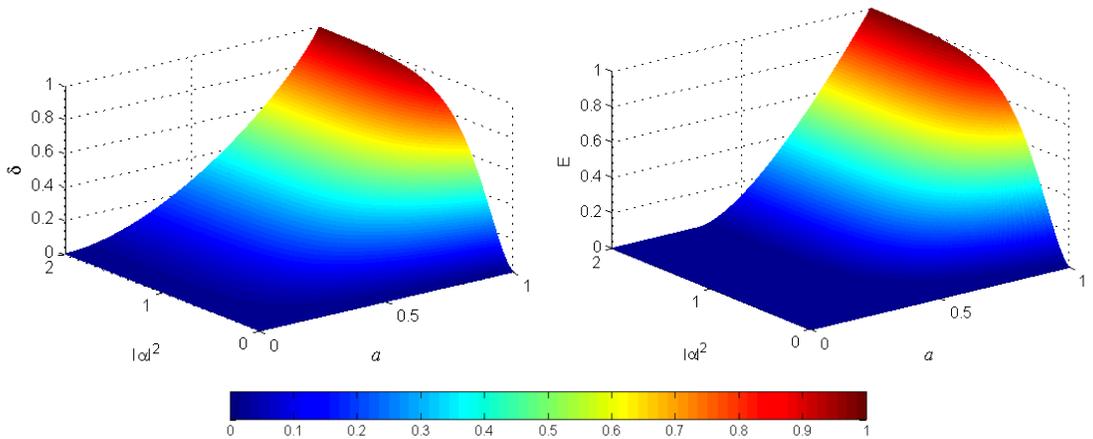

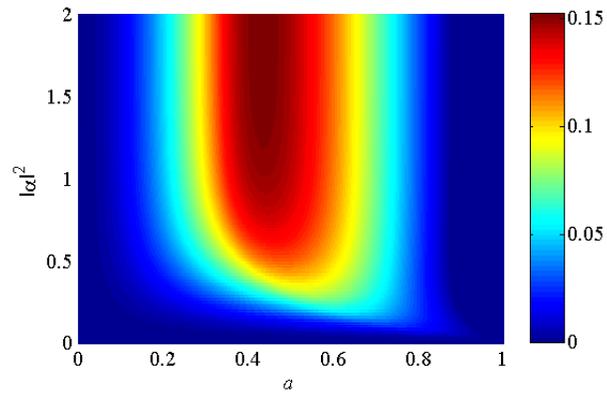

Fig. 4. Variation of Entanglement of formation (*E*), minimum quantum discord ($\delta$) and their difference ($\delta - E$) for quasi-Werner states with respect to mixing parameter '*a*' and mean photon number.